\documentclass[12pt,authoryear]{elsarticle}
\usepackage{graphicx}
\usepackage{float}
\usepackage{newtxtext}
\usepackage{newtxmath}
\usepackage{hyperref}
\usepackage{subcaption}
\usepackage{amsfonts}

\usepackage{dcolumn}
\usepackage{bm}
\usepackage{color}
\usepackage{caption}

\usepackage{amsmath}

\begin{document}

\begin{frontmatter}



\title{On Long-Crested Ocean Rogue Waves Originating From Localized Amplitude and Frequency Modulations}

 \author[label1,label2]{Yuchen He}\ead{yuchen.he@polyu.edu.hk} 
  \author[label3,label4]{Amin Chabchoub}
   \ead{chabchoub.amin.8w@kyoto-u.ac.jp} 
 \affiliation[label1]{organization={Department of Civil and Environmental Engineering, The Hong Kong Polytechnic University},
             city={Hong Kong SAR},
             country={China}}
 \affiliation[label2]{organization={Department of Ocean Science and Engineering, Southern University of Science and Technology},
             city={Shenzhen},
             postcode={518055},
             country={China}} 
\affiliation[label3]{organization={Disaster Prevention Research Institute, Kyoto University},
             city={Uji},
             postcode={611-0011},
             country={Japan}}
\affiliation[label4]{organization={Department of Infrastructure Engineering, The University of Melbourne},
             city={Parkville},
             postcode={3010},
             state={Victoria},
             country={Australia}}

\begin{abstract}
Rogue waves are known to occur on the ocean surface leading to significant damage to marine installations and compromising ship safety. Understanding the physical mechanisms responsible for extreme wave focusing is crucial in order to predict and prevent their formation and impact. Two intensively discussed wave amplification frameworks are the linear and nonlinear focusing mechanisms. These are also known as superposition principle and modulation instability, respectively. We report an experimental study investigating the formation mechanism in a unidirectional representative JONSWAP-type sea state and show that the nonlinear focusing mechanism can be sub-categorized into either a localized amplitude or a so far less-studied phase-related frequency modulation, or both being at play. The frequency modulation-type mechanism occurs at a lower probability, as suggested from the distribution of more than 200 extreme events, however, it cannot be underrated or disregarded. 
\end{abstract}






\end{frontmatter}



\section{Introduction}
Dispersive and nonlinear wave focusing are two fundamental mechanisms responsible for the formation of so-called rogue waves in both finite and infinite water depth when excluding any external influence on the wave field \citep{kharif2008rogue,osbrone2010nonlinear,mori2011estimation,mori2023science,tlidi2022rogue,li2024currents}. While the first process is simply the wave overlap resulting from the dispersive feature of water waves \citep{longuet1957statistical}, the second is based on the modulation instability and disintegration of Stokes waves \citep{benjamin1967disintegration}. Whether the wave overlap or nonlinear focusing is more dominant in the ocean, depends on the steepness, spectral bandwidth, and directionality of the sea state \citep{janssen2004interaction,onorato2006extreme,waseda2009evolution,babanin2011breaking,fujimoto2019impact,dudley2019rogue}. 

The classical nonlinear rogue wave generation mechanism has been so far connected to the long-wave perturbation of a Stokes wave train, and thus, limited to the amplitude modulation (AM) of the waves. This is equivalent to a side-band perturbation in the spectral domain \citep{tulin1999laboratory,zakharov2009modulation}. Another nonlinear focusing mechanism, which has been overlooked and only recently proposed and studied by \cite{sheveleva2020temporal,he2022extreme}, suggests that a localized phase-shift perturbation, or frequency modulation (FM) as we will refer to it in this work, can indeed also generate a strongly localized wave, which satisfies the rogue wave criteria, i.e. $H_{\textnormal{max}}>2 H_S$ or $\eta_{\textnormal{max}}>1.2 H_S$, $H_{\textnormal{max}}$ being the maximal wave height, $\eta_{\textnormal{max}}$ the maximal crest height, and $H_s$ the significant height \cite{kharif2008rogue,mori2023science}. Such FM feature is also included in breather-type solutions of the nonlinear Schrödinger equation (NLSE), which describe the nonlinear stage of the modulation instability process \citep{akhmediev1997solitons,dysthe1999note}, in addition and connection to the wave envelope AM. 

In this work, we provide experimental evidence that a FM-type perturbation is indeed also pertinent for triggering rogue waves in unidirectional random sea states. After running preliminary numerical simulations evolving a narrowband JONSWAP-type wave field, based on the hydrodynamic NLSE to determine and then isolate the extreme events, we then perform water wave tank experiments initiated with the same boundary conditions as the numerical experiments to track the extreme event the flume for the considered time-segment. We then repeat the same experiment, however, commencing with either suppressed phases or suppressed wave amplitude modulations to determine whether the focusing is the result of an initial AM- or FM-type localization or both. The statistical distribution of the 204 extreme events are captured in a histogram and the latter reveals that most extreme events, which occurred as a result of nonlinear wave focusing, are dominantly the result of an AM of a wave field. Nonetheless, the role of initial FM- or both, AM- and FM-type perturbations is still significant and cannot be overlooked.

\section{Numerical Preliminaries}
In order to distinguish and categorize the type of initial focusing conditions in a random wave field, we first perform simplified numerical simulations based on the hydrodynamic NLSE, which is known to be effective in capturing the evolution of extreme waves in narrowband wave processes, despite its simplicity \citep{onorato2013rogue,dudley2014instabilities,chabchoub2016tracking}. We consider a JONSWAP sea state as a reference model for a unidirectional ocean surface \citep{hasselmann1973measurements,janssen2004interaction}. A JONSWAP random wave field satisfies a frequency distribution, parameterized as 
\begin{equation}\label{eqn-JONSWAP spectrum}
    S(f)=\frac{\alpha_P}{f^5}\exp\left[-\frac{5}{4}\left(\frac{f_p}{f}\right)^4\right]\gamma^r,
\end{equation}
where
\begin{equation}
     r=\exp\left[-\frac{(f-f_p)^2}{2\Lambda^2f_p^2}\right],
\end{equation}
$\alpha_P$ is the Phillips constant, determining the spectrum intensity and the significant wave height, $f_p$ is the peak frequency, $\gamma$ is the peak enhancement factor controlling the bandwidth of the spectrum. The constant $\Lambda=0.07$ for $f \leq  f_p$, and $\Lambda=0.09$ for $f > f_p$. The water surface elevation with random phases $\varphi_n$, which are uniformly distributed on $(0,2\pi)$, is then determined by 
\begin{eqnarray}
\eta_{\textnormal{\tiny JONSWAP}}(0,t)&=\sum\limits_{n=1}^{N}{\sqrt{2S\left(f_n\right)\Delta f_n}\cos\left(2\pi f_n t-\varphi_n\right)}.
\label{JONSWAP_TS}
\end{eqnarray} 

Having determined an initially linear JONSWAP-type water surface time-series (\ref{JONSWAP_TS}), which comprises random phases and amplitude variations, we can compute the wave envelope using the Hilbert transform, which we will notate by $\mathcal{H}$, as the following \citep{osbrone2010nonlinear}
\begin{eqnarray}
    \psi_{\textnormal{JONSWAP}}(0,t)&=\eta_{\textnormal{JONSWAP}}(0,t)+i\mathcal{H}\left[\eta_{\textnormal{JONSWAP}}(0,t)\right]\\
        &=A_{\textnormal{JONSWAP}}(0,t)\exp{\left[i\phi\left(0,t\right)\right]}. 
\end{eqnarray}
Here, $\phi(0,t)$ carries the initial phase information. We consider 20.000 waves and choose the wave field to have a representative frequency $f_p=1.2$ Hz, a significant wave height $H_s=0.06$ m while a peakedness parameter is set to be $\gamma=6$. For simplicity, we chose $x=0$ to be our start location for NLSE simulations. This will also be adopted in the experiments. The temporal evolution of the complex envelope can be extracted from the surface elevation time-series using the Hilbert transform \cite{osbrone2010nonlinear,chabchoub2016tracking}. 

\begin{figure}[H]
    \centering
\includegraphics[width=0.88\textwidth]{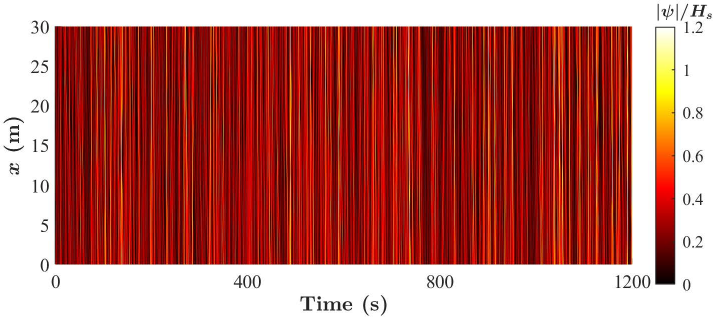}
    \caption{Exemplary spatial evolution of the modulus of wave envelope $|\psi(x,t)|$ for 1200 s and over a spatial distance of 30 m, as simulated by the NLSE (\ref{eqn-time-NLSE}), integrted using the pseudo-spectral method. }
    \label{fig1}
\end{figure}

As previously mentioned, the hydrodynamic time-like deep-water NLSE evolution equation will be used to simulate the evolution of the JONSWAP $\psi_{\textnormal{JONSWAP}}(x,t)$ wave field envelope along the spatial $x$-coordinate  \citep{zakharov1968stability,osbrone2010nonlinear}. It reads 
\begin{equation}\label{eqn-time-NLSE}
    i(\psi_x+\frac{1}{c_g}\psi_t)+\frac{\mu}{c_g^3}\psi_{tt}+\frac{\nu}{c_g}|\psi|^2\psi=0,
\end{equation}
where
\begin{equation}\label{space-NLSE parameters}
    c_g=\frac{\omega_p}{2k_p},\ \mu=-\frac{\omega_p}{8k_p^2},\ \nu=-\frac{1}{2}\omega_p k_p^2,\ \omega_p=\sqrt{gk_p}.
\end{equation}
Here, $g=9.81$ m/s$^2$ is the gravitational acceleration, $c_g$ the group velocity, $k_p$ and $\omega_p=2\pi f_p$ in the above denote the peak wavenumber and angular frequency, respectively, and  $\psi(x,t)$ is the unidirectional complex wave envelope. An example of the numerical spatial evolution of a JONSWAP wave envelope time-series in space is shown in Figure \ref{fig1}
These preliminary numerical simulation results allow us to identify the rogue wave events emerging from the pre-defined random boundary condition. This enables to truncate the later to devote the laboratory efforts only to the wave evolution cases, which guarantee the tracking of extreme events. 

\section{Suppression of the Phase and Amplitude Modulation in the Boundary Conditions} 
Breather solutions of the NLSE are pulsating wave envelopes, which describe local characteristic amplitude and phase modulation dynamics in time and space \citep{akhmediev1997solitons,dysthe1999note}. Recent experimental studies investigated the changes in the breather hydrodynamics when suppressing either one of the two quantities \citep{chabchoub2020phase,he2022extreme}. In fact, the respective breather-like wave fields investigated were found to evolve similarly to the pure breather dynamics, however, with a distinctive delay in the focusing in the propagation direction. 

Following the above-mentioned investigations, we generalize the problem by applying a similar approach to determine the origin of extreme waves in irregular and random sea states, such as in the JONSWAP time-series. Next we describe the modifications in the boundary conditions by suppressing either the phases while keeping the amplitude modulations (AM-only perturbation) or the amplitude modulations while keeping the phases (FM-only perturbation).   

To first-order of approximation, the surface elevation with suppressed phases for pure AM-type boundary conditions and the case of suppressed modulations while retaining the phases are defined as
\begin{equation}\label{eqn-envelope to elevation}
    \eta_{\textnormal{AM/FM}}(x,t)=\frac{1}{2}\left(\psi_{\textnormal{AM/FM}}(x_0,t)\exp[i(k_p x_0-\omega_p t)]+c.c.\right),
\end{equation}
where $c.c.$ denotes the complex conjugate, 
\begin{equation}\label{eqn-AM and FM envelope formulation}
\begin{split}
&\psi_{AM}(0,t)=\left|\psi_{\textnormal{JONSWAP}}(0,t)\right|,\ \textnormal{and}\\
&\psi_{FM}(0,t)=a_c\exp{\left[i\phi\left(0,t\right)\right]},\\
\end{split}
\end{equation} while $a_c=\sqrt{2}\sigma$ being the characteristic wave amplitude and $\sigma$ the standard deviation of the wave field. These latter expressions can be used to define the boundary conditions and to analyze whether the extreme waves emerge as a result of localized AM-type or FM-type perturbation, or influenced by both.

\section{Numerical and Laboratory Experiments}
While we adopted the pseudo-spectral method to integrate the NLSE, the respective laboratory experiments were conducted in a state-of-the-art 30 m water wave tank. Details on the numerical scheme and experimental apparatus can be found in \citet{he2022galilean}.  The numerical and laboratory evolution of the truncated JONSWAP-type wave envelopes are depicted from a propagation distance of 3.75 m until 21.25 m from the wave maker (positioned at $x=0$), which correspond to the first and last wave gauge location, respectively. Moreover, the time-series are shifted by the value of the group velocity $c_g$ for a better display of the focused waves group. The 204 rogue waves, which have been preliminary identified in the numerical hydrodynamic NLSE simulations, are then generated in the wave tank using the same boundary conditions as in the numerical runs one by one. Subsequently, we first suppress phases (AM-only initialization) and in the following the envelope undulations (FM-only initialization), as described in the previous Section. As the number of extreme events is significantly large, we limit our illustrations to a few cases only to emphasize the proof of concept. 
By examining whether an AM or FM initial random localization led to an extreme event in the random JONSWAP wave field, we are able to classify the 204 nonlinear rogue waves initializations into AM-type focusing only, FM-type focusing only, or both combined. 

\begin{figure}[htp]
    \centering
    \includegraphics[width=\textwidth]{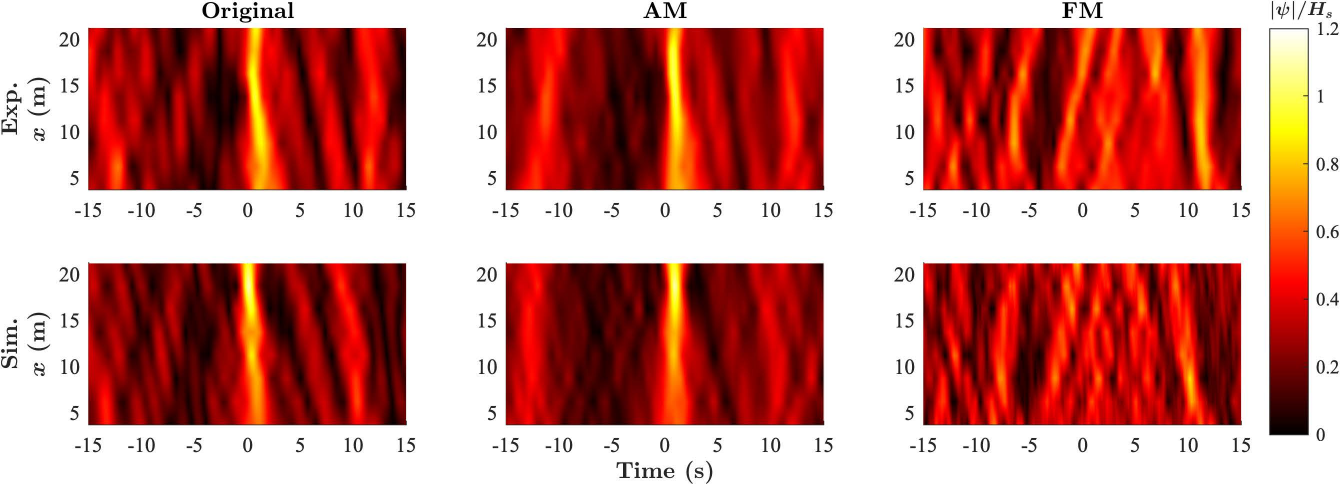}
    \caption{(Rogue Wave No. 47) Modulus of the wave envelope normalized by the value of significant wave height. Left: Original data without any isolation of either phase of amplitude modulation. Middle: Cases of suppressed phases. Right: Cases of suppressed amplitude modulations. Top: Laboratory results. Bottom: NLSE simulation results.}
    \label{fig:AM RW3}
\end{figure}

We start by showing a case for which the AM-type focusing has been identified to be the dominant mechanism in the formation of strong wave focusing. It is shown in Figure \ref{fig:AM RW3}. 



Indeed, we can observe from both laboratory and numerical experiments that the rogue wave event formed in the random JONSWAP wave field can be distinctively explained as being the result of an initial and random AM-type perturbation since removing the initial undulations in the wave envelope while retaining all phases in the same initial time-series does not lead to any extreme wave generation.



\begin{figure}[htp]
    \centering
    \includegraphics[width=\textwidth]{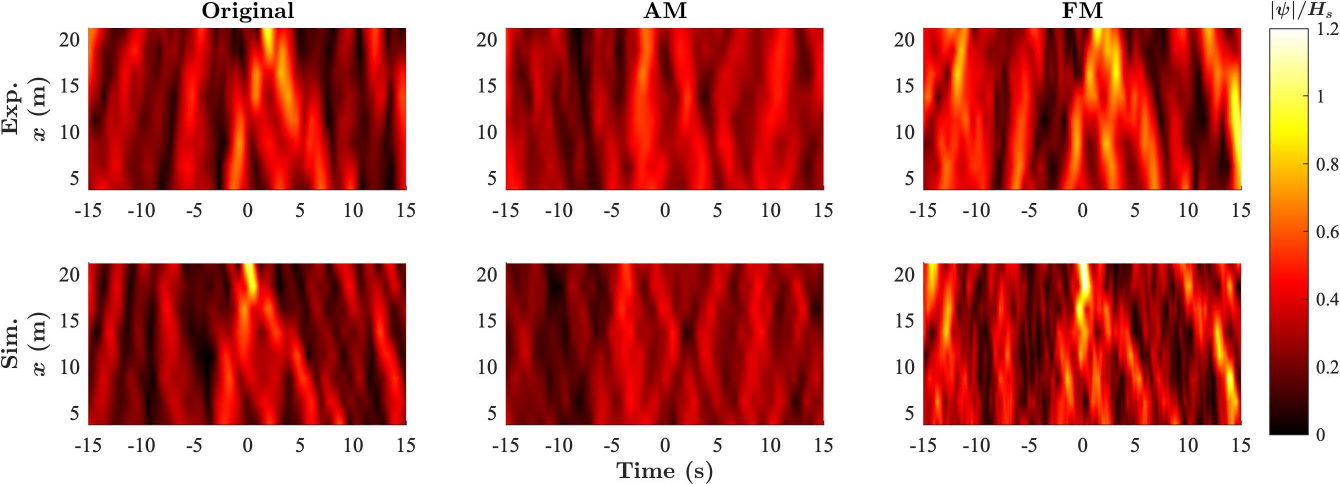}
    \caption{(Rogue Wave No. 167) Modulus of the wave envelope normalized by the value of significant wave height. Left: Original data without any isolation of either phase of amplitude modulation. Middle: Cases of suppressed phases. Right: Cases of suppressed amplitude modulations. Top: Laboratory results. Bottom: NLSE simulation results.}
    \label{fig:FM RW3}
\end{figure}

Many opposite cases have also been identified. One example is shown in Figure \ref{fig:FM RW3}. 

In the latter, we can clearly notice that when suppressing all phases in the wave train, i.e., considering the AM-type initialization approach, a rogue wave generation cannot be observed, neither in the NLSE simulations nor the wave tank data. The focusing appears only when starting the experiments from FM-type boundary conditions with suppressed amplitude modulations.

\begin{figure}[H]
    \centering
    \includegraphics[width=\textwidth]{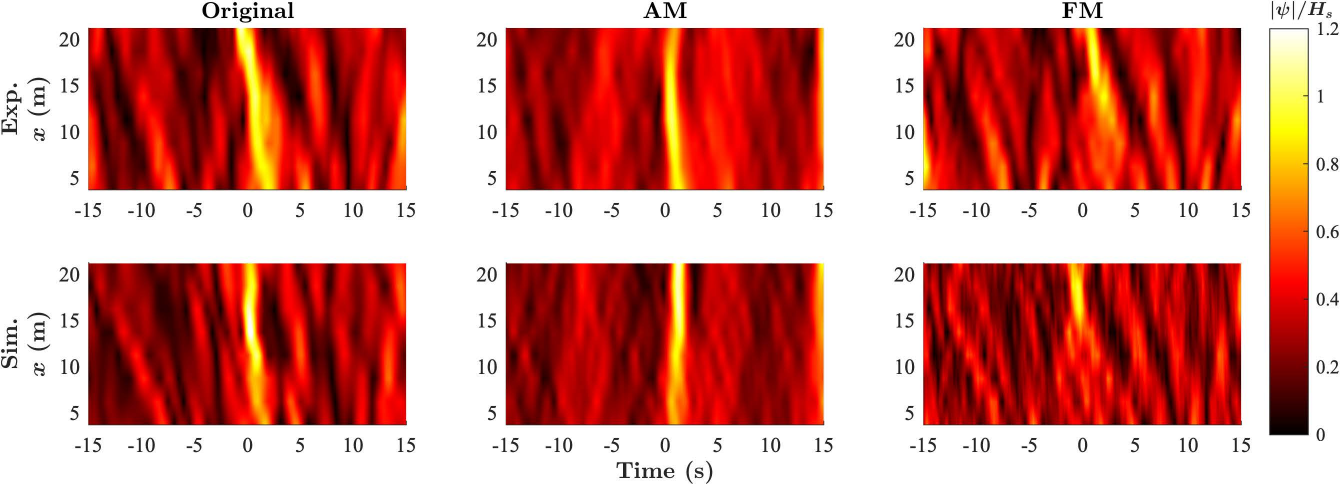}
    \caption{(Rogue Wave No. 185) Modulus of the wave envelope normalized by the value of significant wave height. Left: Original data without any isolation of either phase of amplitude modulation. Middle: Cases of suppressed phases. Right: Cases of suppressed amplitude modulations. Top: Laboratory results. Bottom: NLSE simulation results.}
    \label{fig:mixed RW1}
\end{figure}

Figure \ref{fig:mixed RW1} depicts a case, which highlights the relevance of random AM-type {\it and} FM-type perturbations in the generation of the rogue waves. 

This is indeed the feature of pure NLSE breathers \citep{chabchoub2011rogue,shemer2013peregrine} and an indication that these are embedded in this random wave field \cite{onorato2001freak,osbrone2010nonlinear,chabchoub2016tracking,osborne2020nonlinear,tikan2022prediction}. We will discuss the dominance of each of the two mechanisms or the combination of both in the next Section. 

\begin{figure}[pht]
    \centering
    \includegraphics[width=\textwidth]{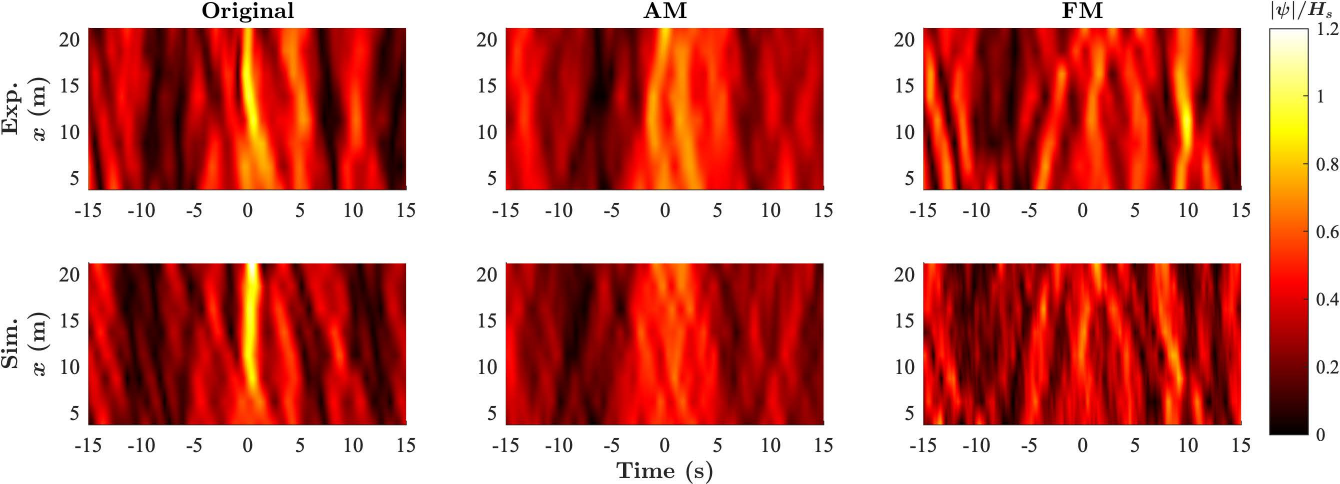}
    \caption{(Rogue Wave No. 26) Modulus of the wave envelope normalized by the value of significant wave height. Left: Original data without any isolation of either phase of amplitude modulation. Middle: Cases of suppressed phases. Right: Cases of suppressed amplitude modulations. Top: Laboratory results. Bottom: NLSE simulation results.}
    \label{fig:mixed RW4-special}
\end{figure}

Note that we also found one extreme wave case for which neither the AM nor FM focusing can explain the extreme wave generation, see Figure \ref{fig:mixed RW4-special}.

We conjecture that this event is a result of wave overlap, which is a principle that should not be ignored when investigating ocean rogue waves \citep{fedele2016real,mcallister2019laboratory,hafner2023machine}. We stress that according to the Rayleigh distribution of linear ocean waves, an extreme event occurs at low probability, which is around 1 in 10.000 waves \citep{kharif2008rogue,mori2023science}. Therefore, considering that we have tested 20.000 waves in total, this isolated case detected is conceptional.

\section{Qualitative Nonlinear Rogue Wave Categorization} 
\begin{figure}[pth]
    \centering
\includegraphics[width=0.75\textwidth]{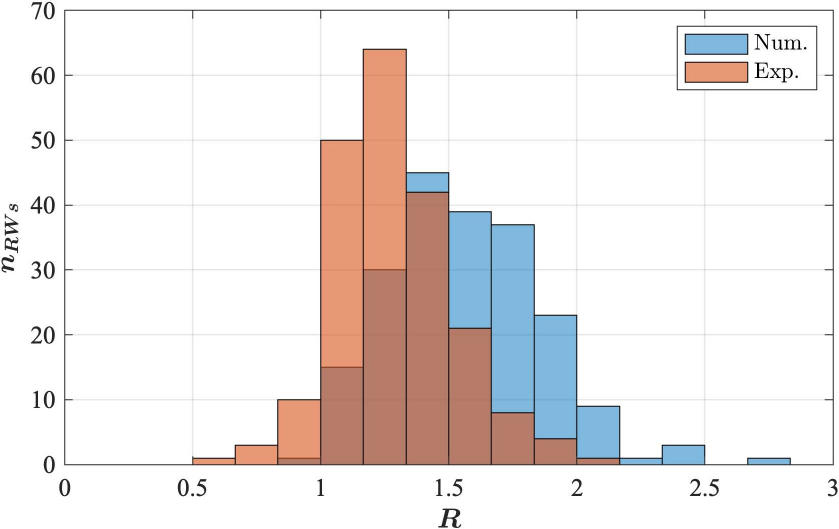}
    \caption{Distribution of the weighting factor $R$, defined by the ratio of the maximum values of the modulus of wave envelopes, see Eq. (\ref{eqn-R factor}), within  10 seconds neighborhood around the extreme events' focal point and obtained from the initial random AM- and FM-only perturbations of the wave field in the numerical (blue) or laboratory (orange) experiments. A relatively higher R values from 1 suggests that AM-type focusing is more dominating, while lower R values suggest that the FM-type focusing is more dominant in triggering the respective wave focusing process.}
    \label{dist}
\end{figure}

In order to quantify the dominance of either the AM-type or FM-type focusing, we introduce a weighting factor $R$ for each one of the 204 rogue wave events. The latter can be simply defined by 
\begin{equation}\label{eqn-R factor}
    R=\frac{\textnormal{max}(|\psi_{AM}(x,t)|)}{\textnormal{max}(|\psi_{FM}(x,t)|)}. 
\end{equation}
We then analyze the distribution of the 204 recorded rogue wave events after applying the latter factor $R$ to our numerical and experimental data. The results are represented in an histogram, see Figure \ref{dist}.
Both experimental and numerical tests reveal in general that most recorded rogue waves are influenced by an AM-type focusing, while the FM-type only mechanism is rather rare, however, not dismissive. The extreme left and right tails of the distribution refer to these cases. Furthermore, the overall shape of the numerical and experimental distributions exhibit a single-peaked quasi-Gaußian shape around 1. 
This suggests that most rogue waves evolve due to a combination of an AM-type and FM-type perturbation. The extreme left and right tails of the distribution refer to the cases, which indicate the potential presence of predominantly either AM-type or FM-type extreme waves.


We emphasize that fewer FM-type rogue waves are observed in the numerical simulations compared to the wave tank experiments. This discrepancy may be attributed to wave breaking or the intrinsic effects of high-order dispersion and nonlinearity, not sufficiently captured by the  simulation framework, based on the hydrodynamic NLSE. 

\section{Conclusion}
We reported an experimental study to distinguish the dominant nonlinear wave focusing subcategories in the formation of long-crested rogue waves in random JONSWAP-type reference sea state. While the wave superposition is a well-established principle for extreme wave generation, the nonlinear focusing can be triggered either from a localized AM, FM, or both. Phase localizations, which are FM-type rogue wave triggers,  have been so far overlooked and missed and even though the resulting nonlinear extreme events being more rare than the ones triggered from AMs, the importance of such a mechanism cannot be underestimated. Overall, the results of hydrodynamic NLSE simulations and laboratory experiments show a reasonable good agreement for all freak wave cases considered, including the ones with initial phase or amplitude undulation suppression in the wave envelope. Future work will focus on studying the relevance of both AM- and FM-type nonlinear focusing mechanisms in directional seas, i.e., short-crested waves. Moreover, supplementary experiments will include a substantial number of added waves. This permits to derive subsidiary quantitative rogue waves categorization criteria when analyzing the respective statistics of extreme events, including the ones generated as a result of dispersive focusing.

\bibliographystyle{elsarticle-harv} 
\bibliography{Refs}

\end{document}